\documentclass[aps,prl,twocolumn,superscriptaddress,nofootinbib]{revtex4-2}

\usepackage{amsmath,amssymb,amsthm,graphicx}
\usepackage{subcaption}
\usepackage[dvipsnames]{xcolor}
\usepackage{multirow}
\usepackage{adjustbox}
\usepackage{comment}
\usepackage{ulem}
\usepackage{enumerate}
\usepackage{hyperref}
\usepackage{cleveref}
\usepackage{bm}

\def\*{\hspace{-1pt}}
\def\newpar{\vskip4pt}

\newcommand{\E}{\mathcal{E}}
\newcommand{\gij}{g_{ij}}

\begin{document}

\title{Optimal Transport Event Representation for Anomaly Detection}

\author{Tianji Cai}
\thanks{Contact author: \href{mailto:tianji\_cai@tongji.edu.cn}{tianji\_cai@tongji.edu.cn}}
\affiliation{School of Physical Science and Engineering, Tongji University, Shanghai 200092, China}
\affiliation{State Key Laboratory of Autonomous Intelligent Unmanned Systems, MOE Frontiers Science Center for Intelligent Autonomous Systems, Tongji University, Shanghai 200092, China}
\author{Aditya Bhargava}
\affiliation{Department of Physics, University of California, Berkeley, CA 94720, USA}
\author{Benjamin Nachman}
\affiliation{Department of Particle Physics and Astrophysics, Stanford University, Stanford, CA 94305, USA}
\affiliation{Fundamental Physics Directorate, SLAC National Accelerator Laboratory, Menlo Park, CA 94025, USA}

\date{\today}

\begin{abstract}
We introduce optimal transport (OT) as a physics-based intermediate event representation for weakly supervised anomaly detection. With only $0.5\%$ injection of resonant signals in the LHC Olympics benchmark datasets, the OT-augmented feature set achieves nearly twice the significance improvement of the standard high-level observables using an idealized setup, while end-to-end deep learning on low-level four-momenta is less effective in this low-signal regime. The observed gains persist across signal types and classifiers considered in this study, suggesting that structured, physics-informed representations can provide a useful complement to existing approaches for anomaly detection.

\end{abstract}

\maketitle

\section{Introduction}

As targeted searches struggle to find new physics beyond the Standard Model, alternative model-agnostic strategies have become increasingly important. Anomaly detection (AD), powered by modern machine learning (ML), offers a promising route to identify rare or unexpected signals without specific model hypotheses~\cite{Kasieczka_2021, Aarrestad:2021oeb, Karagiorgi2021machinelearningsearchnew, Belis_2024}. A prominent class of AD frameworks follows the \textit{weak supervision} (WS) paradigm~\cite{HERNANDEZGONZALEZ201649, Metodiev:2017vrx, PhysRevLett.121.241803, D_Agnolo_2019}, where classifiers are trained to distinguish between signal-rich data and background-only references. These approaches are particularly effective for resonant anomalies, i.e., localized signals atop smooth background distributions, and have been successfully deployed to experimental data~\cite{ATLAS_expt_2020, 2yq5-vj59, CMS_ropp_2025, vvv3-5kkl, via1.0, via2.0, via3.0, gaia_mnras, gaia_curtains}. 

Methods within the WS paradigm differ primarily in background modeling and classifier design~\cite{Hallin_2022, PhysRevD.104.035003, Buhmann:2023acn, Nachman_2020, stein2020unsupervised, cheng2025paws}, both of which depend on the underlying event representation. Most studies use a number of standard high-level observables to summarize the event, which may restrict the scope of detectable signals if the chosen observables cannot capture signal characteristics. To broaden sensitivity, recent works~\cite{Buhmann:2023acn, omnilearn} used as inputs a maximal feature set, i.e., the final-state kinematics of an event. While powerful, methods based on low-level four-momenta usually require training (or pre-training~\cite{omnilearn}) a large (foundation) model on substantial amounts of data, and are typically less effective in the (ultra) low-signal regime where anomaly detection is most needed.

We propose an alternative approach that exploits full final-state kinematics while avoiding sophisticated, resource-intensive models and extensive training data. In particular, \textit{optimal transport} (OT) theory is introduced to define a novel intermediate representation, capturing essential event structures in a compact and structured form. OT defines a distance between probability distributions by computing the most efficient way to transform one into another~\cite{computationalotbook}. Its sensitivity to geometric structure has been utilized to equip the space of collider events with physically meaningful metrics~\cite{Komiske:2019fks, Cai:2023edb, PhysRevD.102.116019, PhysRevD.105.076003, cai2024phasespacedistancecollider, Komiske:2020qhg, Larkoski:2023qnv, Gambhir:2024ndc}, showing promises across collider applications including jet and event classification~\cite{Onyisi_2023, cai2025multiscaleoptimaltransportcomplete}, calibration~\cite{ATLAS:2025rbr}, and more recently, anomaly detection at the event level~\cite{Craig:2024rlv, brennan_cai_2025, dagnolo2025intrinsicdimensioncolliderevents}.

Nonetheless, the full potential of optimal transport as an \textit{event representation} remains largely untapped. Here we propose the conceptual shift: rather than merely using OT to define metrics between events, we develop an efficient representation through its linearization. When combined with standard high-level observables, features derived from this OT representation yield noticeable improvements for resonant anomaly detection, particularly in the low-signal regime. Notably, only a small number of OT-based features is required to enhance performance, with further gains achieved when augmented by additional higher-order subjettiness. These results highlight the value of physically grounded representations as a bridge between engineered features and fully end-to-end machine learning, encouraging further exploration of model latent spaces and alternative intermediate representations beyond the specific constructions considered here.

\section{Datasets and Weak Supervision Framework}

We use datasets from the 2020 LHC Olympics (LHCO) Anomaly Detection Challenge~\cite{Kasieczka_2021, gregor_kasieczka_2022_6466204, david_shih_2023_8370758}, which includes two signal-injected sets (R\&D1 and R\&D2) and a background-only set. Each R\&D dataset contains one million QCD dijet events and $100$ thousand signal events from a resonant decay $W' \to XY$ with $m_{W'}, m_{X}, m_{Y}=3.5\text{ TeV}, 500\text{ GeV}, 100 \text{ GeV}$, respectively. In R\&D1, $X, Y \to qq$, producing two-pronged jets; in R\&D2, $X, Y \to qqq$, yielding three-prong substructure.

Events are simulated in \texttt{Pythia 8}~\cite{Pythia, Pythia_manual} and \texttt{Delphes 3.4.1}~\cite{Delphes_intro}, whereas jets are clustered in \texttt{Fastjet}~\cite{Cacciari_2012} via the anti-$k_T$ algorithm~\cite{Cacciari_2008} with $R = 1$. Events whose leading jet has $p_T > 1.2$ TeV are selected. Only particles in the two leading-$p_T$ jets are retained, which are expected to contain the $X/Y$ decay products. The signal region (SR) is defined as $m_{JJ} \in [3.3, 3.7] \text{ TeV}$, with its complement forming the sideband regions (SB). This is chosen for consistency with prior works~\cite{Hallin_2022, Buhmann:2023acn}. Each R\&D dataset is further augmented with $\sim$610k background events in the SR from Ref.~\cite{david_shih_2023_8370758}, yielding $\sim$731k background and $\sim$75k signal events in total.

Weak supervision trains on mixed samples without requiring event-level labels or class proportions, and is provably as powerful as full supervision by the Neyman–Pearson lemma in asymptotic limit~\cite{doi:10.1098/rsta.1933.0009}. This underlies the classification without labels (CWoLa) framework~\cite{Metodiev:2017vrx}, where recent CWoLa-based anomaly detectors model SR background by interpolating SB data via classifiers~\cite{PhysRevD.101.095004}, density estimation~\cite{Nachman_2020, stein2020unsupervised}, generative models~\cite{Hallin_2022}, and invertible networks~\cite{Raine:2022hht}. These methods are now capable of near optimal matching. We therefore assume perfect background interpolation and adopt an \textit{idealized anomaly detector} (IAD)~\cite{Hallin_2022}, which enables us to isolate the effect of the proposed representation and to perform a controlled representation-level comparison within the weakly supervised setting. \footnote{The IAD framework should not necessarily be interpreted as a strict upper bound on anomaly detection performance. In realistic weakly supervised analyses, background mismodeling may either reduce or, in some circumstances, artificially increase the apparent discrimination power of a classifier.} Since background-modeling effects are removed by construction, the resulting performance primarily reflects the discriminating power of the underlying representation itself, rather than the details of a particular background-estimation strategy. The IAD framework is therefore well suited to the central goal of this work, namely understanding the physical information encoded in the novel OT event representation.

Two training sets (A1 and A2) are constructed in the signal region. The set A1 contains only background events (labeled 0), whereas the set A2 includes background plus a fixed fraction of signal events, i.e., S/B=$0.5\%, 0.63\%, 0.7\%, 0.8\%, 1\%, 1.2\%, 1.4\%, 1.6\%, 10\%$, covering a large spectrum of possible signal amounts. In total, there are nine A2 sets, one for each S/B level, and all events in A2 are labeled 1. Classifiers are then trained to distinguish events in A1 from those in A2, using only dataset-level labels. The main results use boosted decision trees (BDTs)~\cite{finke2023BackToTheRoots, bdt-intro-aux} as the classifier, thanks to its robustness to noise and correlated features. Results using multilayer perceptrons (MLPs) are also presented in the Appendix for comparison. To reduce statistical fluctuations, we aggregate over 50 independently trained BDT classifiers (10 for MLPs as in~\cite{Hallin_2022}) to obtain the results for one ensemble, and further average over 10 ensembles to estimate the mean and the spread.\footnote{A tree consists of 200 iterations with up to 31 leaf nodes. An MLP has three hidden layers, each with 64 ReLU units. We found the results to be robust to reasonable architectural variations and therefore did not further optimize model hyperparameters.} Performance is evaluated using event-level labels on two separate test sets---background-only B1 and signal-only B2. On the statistical level, our datasets match those used in Ref.~\cite{Hallin_2022}, with A1 having $272$k background events, A2 containing 121,352 background events plus injected signal events, B1 having 340k background events, and B2 having 20k signal events---all within the SR.

\section{Optimal Transport Representation and Derived Features}

Standard analyses rely on high-level observables such as jet mass and $n$-subjettiness ratios~\cite{Thaler:2010tr, Thaler:2011gf}. The official LHCO R\&D1 itself provides $\{m_{J_1}, m_{J_2}, \tau_{21}^{J_1}, \tau_{21}^{J_2}\}$ for the two leading jets, whereas R\&D2 adds $\{\tau_{32}^{J_1}, \tau_{32}^{J_2}\}$. Here we propose an intermediate representation based on particle four-momenta, using a specific OT distance called the 2-Wasserstein metric (W$_2$). For a pair of events $\E$ and $\Tilde{\E}$ containing massless constituent particles at $x_i, \Tilde{x}_j \in (y, \phi) \subseteq \mathbb{R}^2$ with normalized transverse momenta $p_T^i, \Tilde{p}_T^j$ respectively, the 2-Wasserstein distance is defined as 
\begin{align} \label{eq:Wp}
    \text{W}_2(\E, \Tilde{\E}) & = \min_{\gij\in\Gamma(\E, \Tilde{\E})}\left(\sum_{ij} \gij || x_i -\Tilde{x}_j||^2 \right)^{1/2},\\
    \Gamma(\E, \Tilde{\E}) & =\left\{\gij:\gij\geq 0, \sum_j\gij=p_T^i, 
    \sum_i \gij =\Tilde{p}_T^j\right\}, \nonumber
\end{align}
where $\Gamma(\E, \Tilde{\E})$ contains the transport plans that move $\gij$ amount of $p_T$ from particle $i$ in $\E$ to particle $j$ in $\Tilde{\E}$ over a distance $|| x_i -\Tilde{x}_j||$ on the $y-\phi$ plane. Thus, W$_2$ quantifies the minimal \textit{cost} to morph one event into another and is by definition infrared and collinear (IRC) safe. 

To convert the pairwise distance into an event representation, we employ the linearization method originally introduced to reduce OT's computational cost~\cite{PhysRevD.102.116019, caihkdistance}. The linearized W$_2$ (LinW$_2$) embeds each event into the tangent space at a fixed reference event $\mathcal{R}$ containing particles with $p_T=R_i$ at $y_i \in (y, \phi)$. Its $i$-th component is given by
\begin{equation}\label{eq:LinW2coord}
    \text{LinW$_2$}^i(\E)|_\mathcal{R} := \frac{1}{\sqrt{R_i}} \sum_j r_{ij}x_j \quad \text{for } i \in [1, R],
\end{equation}  
where $r_{ij}$ is an optimal transport plan from $\mathcal{R}$ to $\E$ and the RHS gives the $p_T$-weighted barycenter of where each reference particle is mapped. Since $x_j=(y_j, \phi_j)$, $\text{LinW$_2$}^i$ is itself a vector in $\mathbb{R}^2$, resulting in a $2R$-dimensional embedding for each event which is likewise IRC-safe. Instead of treating LinW$_2$ as merely an efficient approximation to the original W$_2$ distance, we highlight this novel representation in its own right, marking a major conceptual shift from earlier works.

In practice, the LinW$_2$ embedding is applied at the jet level, where each jet is preprocessed by centering its axis and vertically aligning the principal component of its transverse momentum flow. The reference jet $\mathcal{R}$ is constructed as a $10 \times 10$ grid over the $(y,\phi)$ plane, with every particle assigned a uniform $p_T=\frac{1}{100}$.\footnote{The mathematical advantages of using a uniform reference jet were established in~\cite{PhysRevD.102.116019}. The impact of alternative, including non-uniform, references has been analyzed in~\cite{caihkdistance, PhysRevD.105.076003} and found to be minimal provided the reference is neither too concentrated nor too sparse. Accordingly, we adopt a uniform reference with a number of constituents comparable to that of particles in a typical jet.} The W$_2$ distance between this $10 \times 10$ uniform reference and a given jet serves as a measure of the isotropy of the jet radiation pattern~\cite{Cesarotti:2020hwb}, and also approximately gives its $100$-subjettiness value, in accordance with the geometric reformulation of $n$-subjettiness observables in terms of an OT distance~\cite{Komiske:2020qhg}.

To further obtain the representation for an entire event, we concatenate the LinW$_2$ embeddings of the two leading jets, yielding a Euclidean vector in $\mathbb{R}^{2\times100\times2=400}$. Extensions to a more general multiscale OT treatment for an entire collider event~\cite{cai2025multiscaleoptimaltransportcomplete} are left to future work. Constructing the OT event representation scales at most as $\mathcal{O}(n^3)$ in the number of particles in the reference jet $\mathcal{R}$, which is negligible for $n\sim\mathcal{O}(100)$ and is further parallelizable across events under the linearization scheme.

\begin{figure} 
    \centering
    \includegraphics[width=0.75\linewidth]{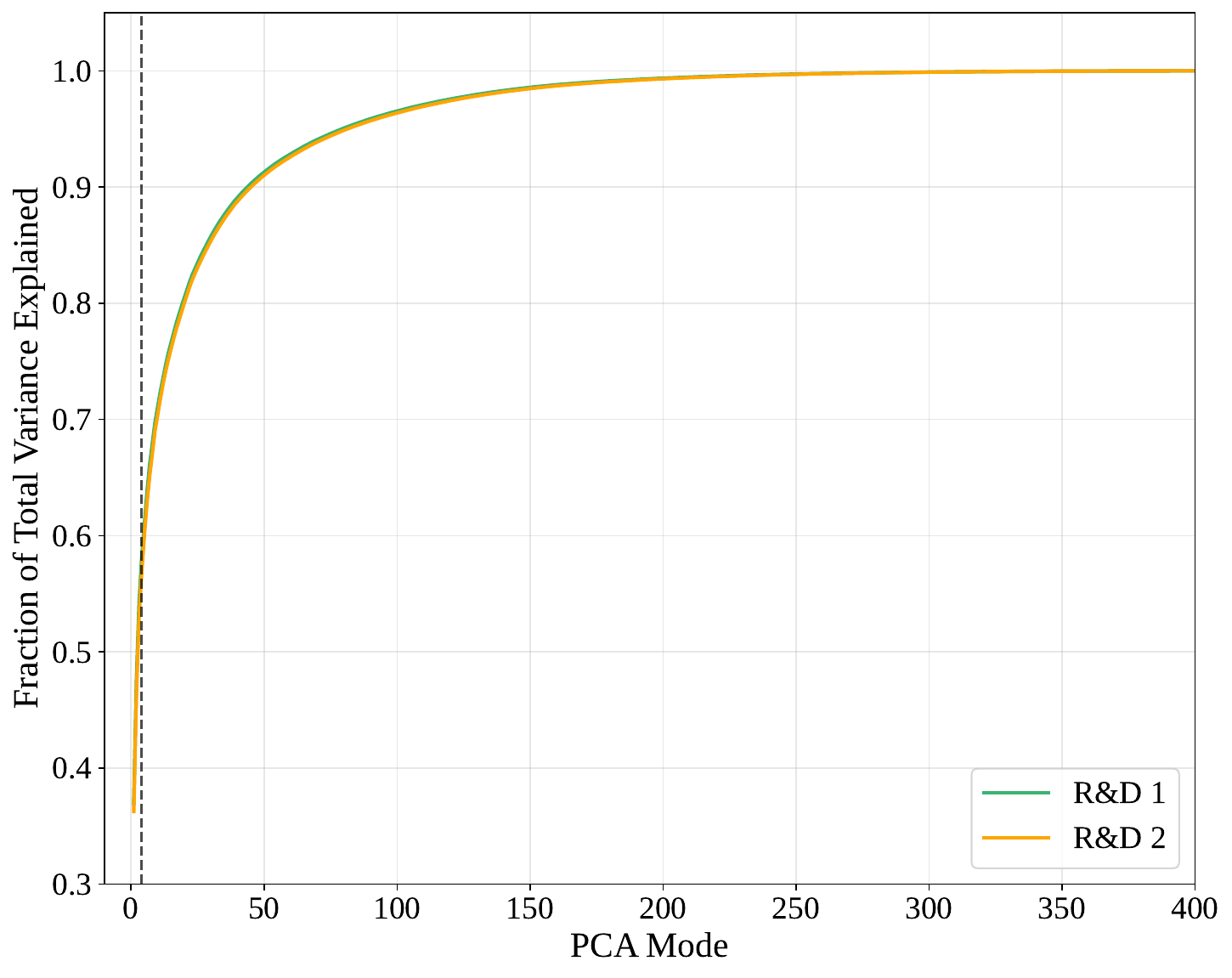}
    \caption{Total variance explained by increasing numbers of PCA modes of the OT representations for 10k samples from the R\&D1 dataset (\textit{green}) and the R\&D2 dataset (\textit{orange}).}
    \label{fig:pca_variance_plot}
\end{figure}

From the high-dimensional OT representation, we further extract features using a simple principal component analysis (PCA). For a diagnostic sample consisting of 5k background and 5k signal events randomly selected from R\&D1 or R\&D2, the leading four modes account for almost $60\%$ of the variance, while approximately $100$ modes suffice to capture more than $95\%$ of total data variance; see Fig.~\ref{fig:pca_variance_plot}. Note that this diagnostic sample serves only to illustrate the variance structure of the OT representation and plays no role in the anomaly detection analysis. Instead,  for each signal injection level in the IAD analysis, the PCA basis is fitted exclusively on the corresponding unlabeled training sample $A_1+A_2$, without access to event-level signal/background labels. The test sets $B_1$ and $B_2$ are subsequently projected onto this training-defined basis, with the first $k$ principal components used as the OT$_k$ feature set.

To further characterize the physical information encoded in the OT representation, we compare the leading OT PCA modes with standard jet observables using Pearson correlations and mutual information, which respectively probe linear and nonlinear relationships; the results are summarized in Table~\ref{tab:ot_correlations} in the Appendix. We find that the OT features are partially correlated with jet masses and higher-order $N$-subjettiness ratios, consistent with the known geometric relation between OT and subjettiness. However, the correlations and mutual information values remain far from saturated, indicating that the OT representation captures overlapping but complementary information beyond that contained in jet mass and subjettiness. This interpretation is further supported by the improved performance observed when OT features are combined with standard observables in the following section.

\section{Results}

\begin{figure*}[htbp]
    \centering
    \includegraphics[width=\linewidth]{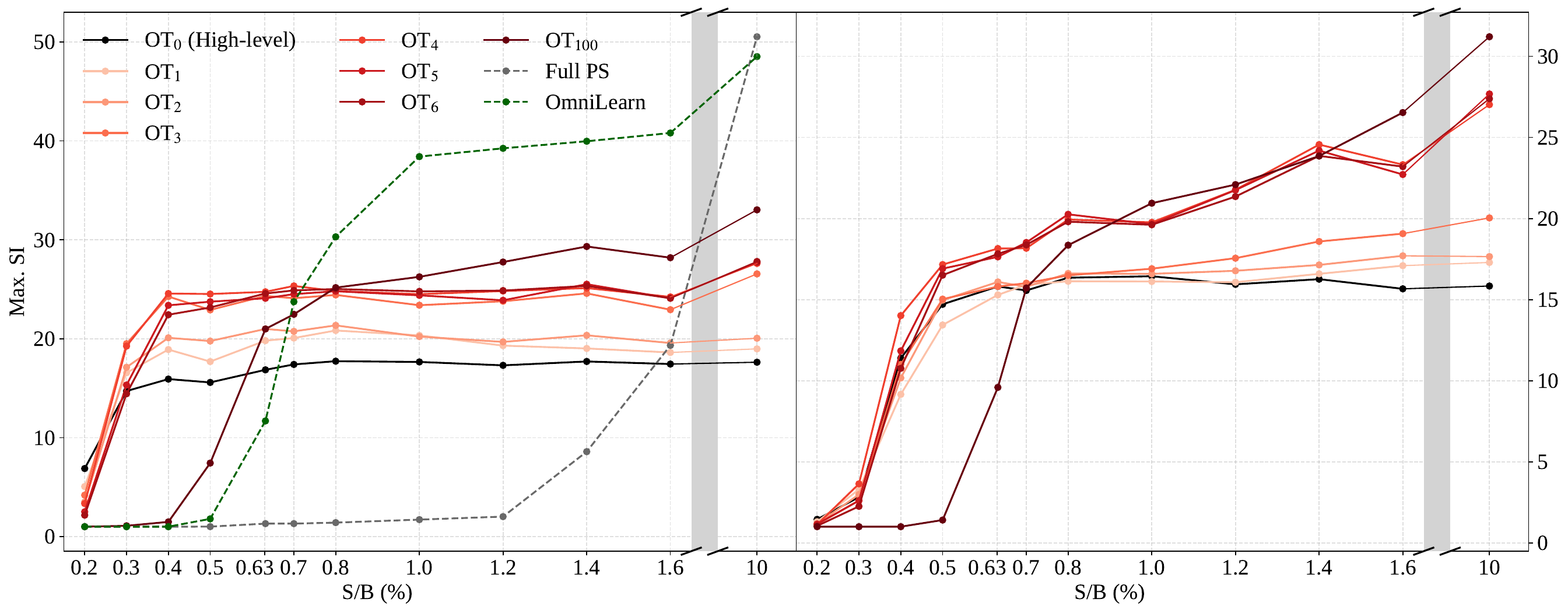}
    \caption{Maximum Significance Improvement (SI) for the R\&D1 (\textit{left}) and R\&D2 (\textit{right}) datasets using BDT ensembles as the classifier, with the signal injection level S/B varying from $0.2\%$ to $10\%$. The group of red curves represent increasing numbers of OT features added to standard high-level observables, i.e.,  $\{m_{J_1}, m_{J_2}, \tau_{21}^{J_1}, \tau_{21}^{J_2} \}$ for R\&D1, and $\{m_{J_1}, m_{J_2}, \tau_{21}^{J_1}, \tau_{21}^{J_2}, \tau_{32}^{J_1}, \tau_{32}^{J_2} \}$ for R\&D2. The dashed gray line in the left subplot (R\&D1) shows the max SI values from Ref.~\cite{Buhmann:2023acn} using full phase space as inputs to dedicated models, whereas the dashed green line shows the performance of the pre-trained foundation model OmniLearn from Ref.~\cite{omnilearn}. Both the full phase space method and foundation models are evaluated in the same IAD setup.}
    \label{fig:vary_sig_bdt}
\end{figure*}

\begin{figure*}[htbp]
    \centering
    \includegraphics[width=\linewidth]{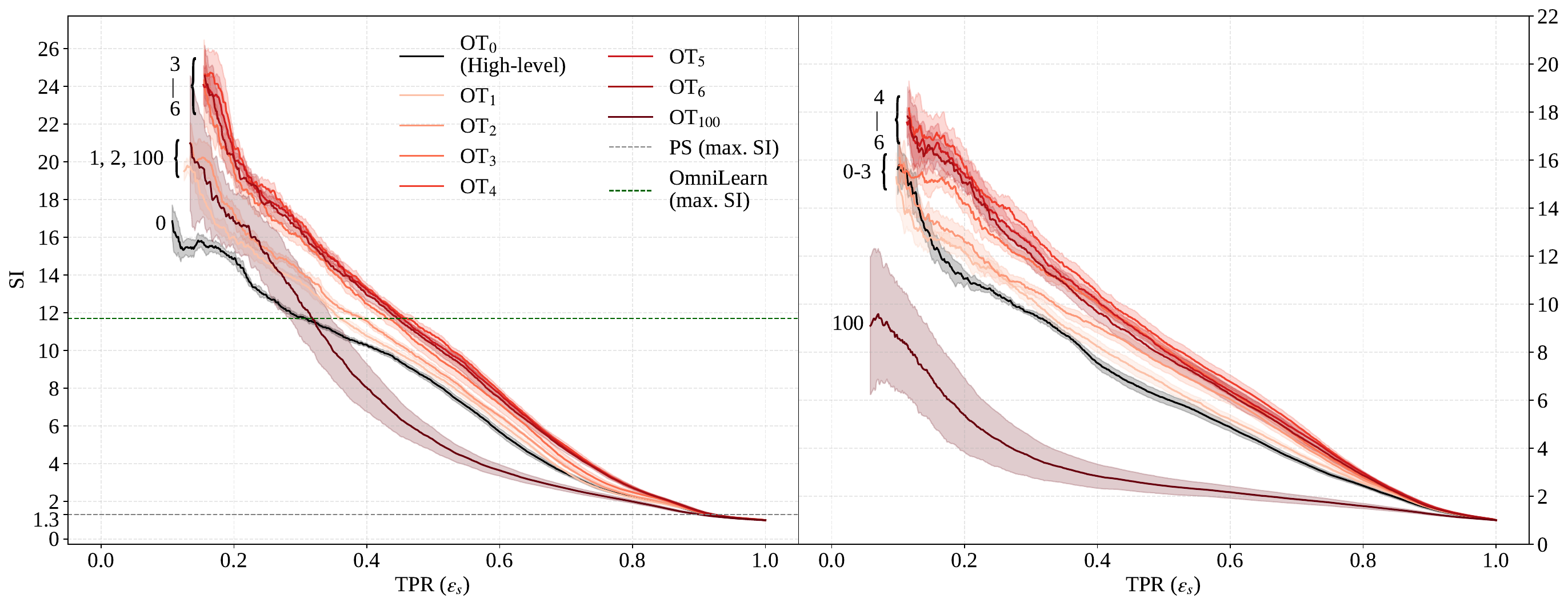}
    \caption{Significance Improvement (SI) curves for the R\&D1 (\textit{left}) and R\&D2 (\textit{right}) datasets at a representative signal injection level of S/B$=0.63\%$, using BDT ensembles as the classifier. The group of red curves represent increasing numbers of OT features added to standard high-level observables, with shaded bands indicating $1\sigma$ variations across the BDT ensembles. For R\&D1, the dashed gray line shows the maximum SI values of the full phase space method from Ref.~\cite{Buhmann:2023acn}, and the dashed green line shows that of the pre-trained foundation model OmniLearn from Ref.~\cite{omnilearn}, where both are evaluated in the same IAD setup.}
    \label{fig:const_sig_bdt}
\end{figure*}

Fig.~\ref{fig:vary_sig_bdt} shows the maximum significance improvement (SI) obtained by scanning the signal-to-background ratio (S/B) from $0.2\%$ to $10\%$ for the R\&D1 and R\&D2 datasets using BDT ensembles as the classifier. At low signal injection levels (S/B $\lesssim 1.6\%$), feature sets augmented with the first $k$ PCA modes of the OT representation (OT$_k$) consistently outperform both standard high-level observables provided in the official LHCO datasets and the low-level full phase space (PS) method of Ref.~\cite{Buhmann:2023acn}. In the ultra-low regime (S/B $\lesssim 0.7\%$), OT$_k$ exceeds the pre-trained foundation model (OmniLearn) of Ref.~\cite{omnilearn}, despite the latter's substantially greater computational cost. Fig.~\ref{fig:const_sig_bdt} further displays the full SI curves at a representative $S/B=0.63\%$.  

On R\&D1, the enhancement is most pronounced in the ultra-low to low signal regime, where approaches based on the full phase space and foundation models both suffer from limited signal statistics, with the latter alleviating the issue to some extent through large-scale pre-training. With less than $0.5\%$ signal injection, OT-augmented feature sets raise the maximum SI to $\gtrsim 25$, more than an order of magnitude above the low-level baselines and $\sim 65\%$ above the standard high-level observable set $\{m_{J_1}, m_{J_2}, \tau_{21}^{J_1}, \tau_{21}^{J_2}\}$. The performance of OT$_k$ (except $k=100$) remains stable across the entire range of signal fractions. Similar gains are observed in the less-studied R\&D2 dataset, albeit with a smaller margin over the standard observables $\{m_{J_1}, m_{J_2}, \tau_{21}^{J_1}, \tau_{21}^{J_2}, \tau_{32}^{J_1}, \tau_{32}^{J_2}\}$. No full phase space or foundation model benchmark currently exists for R\&D2; based on the R\&D1 comparison, comparable trends are likely and we encourage future studies to strengthen the low-level benchmarks.

Only a few OT features are needed to achieve the observed gains. The first 3–5 PCA components already saturate the maximum SI, with negligible benefit beyond $k \approx 6$, consistent with the variance profile in Fig.~\ref{fig:pca_variance_plot}. Using very large OT feature sets (e.g., OT$_{100}$) degrades performance for S/B $\lesssim 0.6\%$, partially reflecting the difficulty of training BDT ensembles with many correlated inputs. The effect is even stronger for MLP classifiers; see Fig.~\ref{fig:vary_sig_R&D1} in the Appendix. This illustrates a fundamental trade-off in anomaly detection: expanding the feature set increases the information accessible to more expressive classifiers, but also requires higher signal statistics for effective training. Consequently, well-designed representations and careful feature selection are essential for maximizing discriminative power and attaining optimal sensitivity, especially in the low-signal regime.

In the high-signal regime ($\sim10\%$), the two low-level approaches perform best, reaching a maximum SI of 50 on R\&D1, consistent with their ability to exploit raw four-momenta when ample signal is present. OT$_{100}$ attains a max SI of 33, $\sim 20\%$ above OT$_{3-6}$ and nearly 80\% above the standard observable set $\{m_{J_1}, m_{J_2}, \tau_{21}^{J_1}, \tau_{21}^{J_2} \}$. An analogous pattern appears in R\&D2 at S/B=10\%, where OT$_{100}$ reaches a max SI $\approx 32$, more than doubling the performance of the corresponding jet mass and $n$-subjettiness ratio inputs. The gap between OT$_{100}$ and the low-level methods likely indicates the presence of information not encoded in the OT representation (see below), possibly associated with IRC-unsafe aspects of the event. Since the OT representation is IRC-safe by construction, its gains are expected to be more robust across signal types and more transferable from simulation to data.

An ablation study on R\&D1 at S/B=0.63\% further isolates the role of OT features; see Fig.~\ref{fig:ablation_R&D1} in the Appendix. OT features alone, without $m_J$ or $\tau_{21}$, do not yield competitive performance. However, including jet masses and at least the first four OT modes raises the maximum SI to $\sim 18$, matching the performance of the standard observable set in Fig.~\ref{fig:vary_sig_bdt}. This confirms that OT features encode similar morphological information as $n$-subjettiness, while the absence of explicit jet-mass information in the current OT construction explains the lower SI ($\sim 12$) obtained when the inputs include only $\tau_{21}$ and OT$_{3-6}$.

On the other hand, to assess whether OT features provide information beyond traditional subjettiness, we repeat the analysis on the R\&D1 dataset over S/B=0.2\%--10\% using BDT ensembles. The inputs now include the \textit{Baseline} and three \textit{Extended Sets} from Ref~\cite{finke2023BackToTheRoots}, consisting of $\tau_n^\beta$ up to $n=9$ and $\beta=0.5, 1, 2$. These input sets are further augmented by the first four PCA modes of the OT representation (OT$_4$), as more OT features are found to be redundant from earlier results. 

\begin{figure*}[htbp]
    \includegraphics[width=\linewidth]{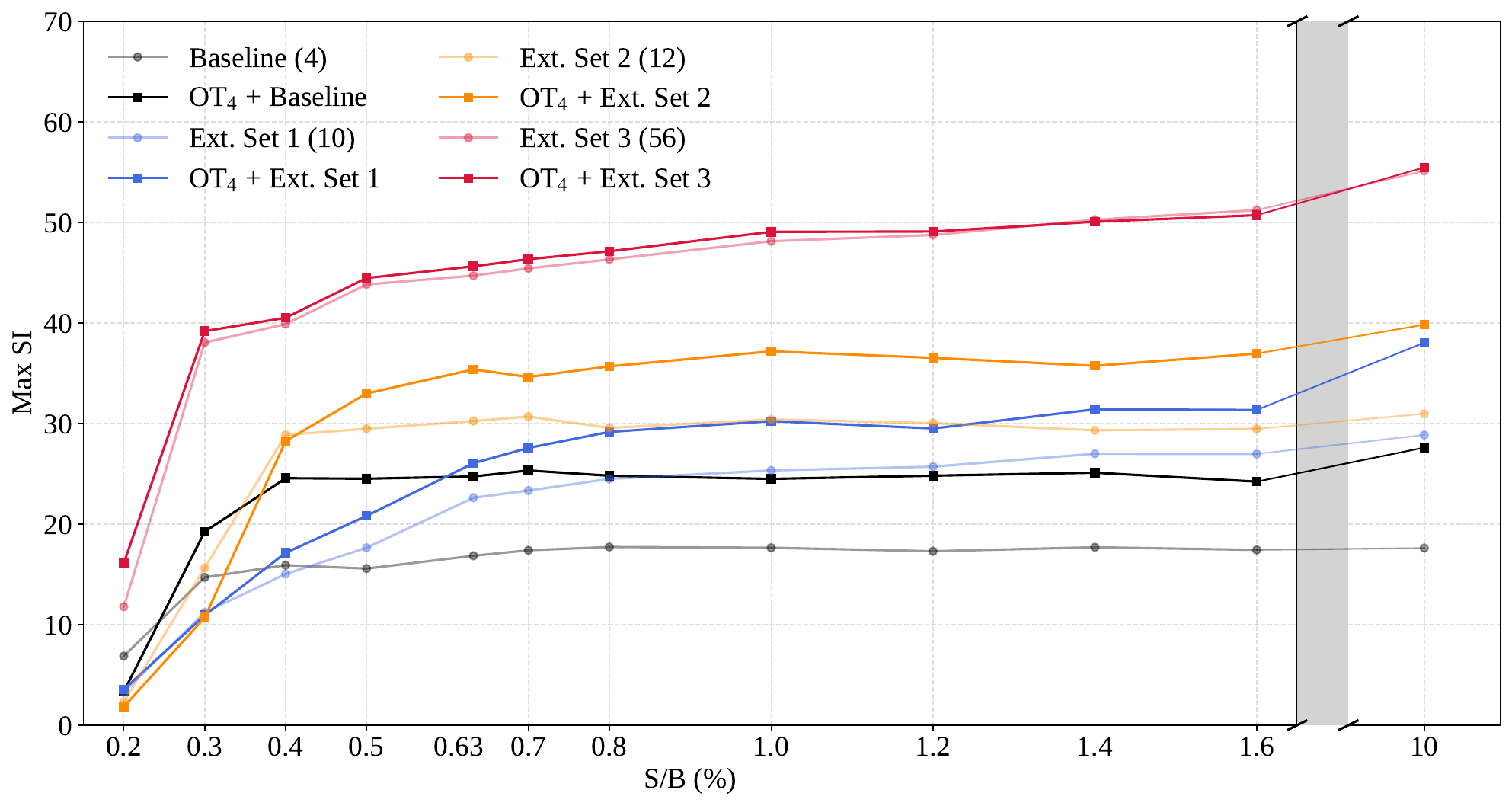}
    \caption{Maximum SI for the R\&D1 dataset using BDT ensembles as the classifier, with the signal injection level S/B varying from $0.2\%$ to $10\%$. The input features consist of the Baseline and three Extended Sets from Ref.~\cite{finke2023BackToTheRoots}, together with the first four PCA modes for the OT representation. Lighter (darker) colors denote feature sets without (with) OT inputs. Numbers in parentheses indicate the dimensionality of each feature set in Ref.~\cite{finke2023BackToTheRoots}.}
    \label{fig:extended_R&D1}
\end{figure*}

As shown in Fig.~\ref{fig:extended_R&D1}, the inclusion of OT features improves performance across all feature sets, indicating that the OT representation does capture complementary information beyond traditional subjettiness observables. For Extended Set 3, the improvement becomes marginal, consistent with the similarity between the current OT linearization scheme and high-order subjettiness (e.g., $\tau_{100}$). In principle, subjettiness observables are encoded in the OT representation itself~\cite{Komiske:2020qhg}; extracting this information in practice, however, requires more expressive feature extraction methods than PCA, and is left for future study.

Interestingly, Extended Set 3 (with OT) achieves a maximum SI of $\sim 45$ at S/B=0.5\% and further exceeds 55 at S/B=10\%, surpassing even the two low-level approaches (both ineffective at S/B=0.5\% and with a max SI $\sim 50$ at S/B=10\%). This suggests that much of the information learned by the full phase space method and foundation models is likely to have already been encoded in the combined subjettiness and OT features, again underscoring the importance of physics-informed representations while motivating further study of what large models capture.

\section{Conclusions}

We presented a physics-aware intermediate representation of collider events based on optimal transport and explored its application to resonant anomaly detection. Within the controlled idealized anomaly detection framework adopted in this work, augmenting standard high-level observables with a small number of OT-derived features leads to substantial improvements in anomaly detection performance, with significance improvements above $25$ at signal injection levels of approximately $0.5\%$ in the LHCO R\&D1 benchmark and comparable gains in R\&D2. Taken together, these results highlight the potential value of physics-informed intermediate representations as a bridge between handcrafted observables and end-to-end machine learning for weakly supervised collider searches.

Several extensions could further enhance performance. While principal component analysis was employed for its simplicity and efficiency, more expressive approaches to feature extraction could be explored. Likewise, advanced classifiers such as neural networks may enable direct use of the complete OT representation without dimensionality reduction. Conversely, the surprising effectiveness of the simple PCA method underscores the importance of carefully balancing data and computational efficiency when introducing sophisticated models, particularly in the low-signal regime.

Beyond boosted dijet signals, our framework is well adapted to more complex event topologies. OT-based observables have demonstrated sensitivity to high-multiplicity, quasi-isotropic radiation patterns typical of hidden-valley scenarios~\cite{Cesarotti:2020hwb, Strassler_2007, Han_2008}. In such new physics settings, OT may be particularly advantageous, as it provides a more flexible and globally structured representation than traditional subjettiness observables. For events featuring distinct total transverse momenta and multiple intrinsic scales, more flexible OT representations with additional degrees of freedom may be required~\cite{cai2025multiscaleoptimaltransportcomplete, caihkdistance}. The framework can also be generalized to incorporate information beyond event kinematics, such as particle ID, through multi-species vector-valued optimal transport~\cite{craig2025vectorvaluedoptimaltransport}. Preliminary studies further suggest applicability of the OT representation to non-resonant anomaly detection, which may open a largely unexplored landscape for new physics discovery.

At the same time, the observed relationship between higher-order subjettiness and OT features underscores the need for a systematic understanding of the physics content encoded in the OT representation. The feature-level diagnostics presented here suggest that OT captures information that is partially overlapping with, yet not fully described by, conventional observables such as jet mass and subjettiness. Since shape-related observables can in principle be derived directly from the OT formulation~\cite{Komiske:2020qhg}, OT may provide a unified geometric representation of collider events from which a broad class of physics observables can be recovered. Understanding the physical structures encoded in such representations across a wider range of signal and background processes, and determining their predictive capacity, therefore represents an important direction for future study.

In conclusion, the strong performance of both the OT representation and the extended feature sets compared to the available low-level approaches in the IAD setting highlights the importance of physics-aware representations for collider anomaly detection. This advantage appears particularly pronounced in the (ultra) low-signal regime, where inductive biases grounded in physical principles may provide higher sensitivity than purely data-driven approaches.

The code for the analysis can be found at \url{https://github.com/TianjiCai/ADwithOT}.

\begin{acknowledgments}
\newpar
\paragraph{Acknowledgments}---\,
The authors would like to thank Michael Krämer and Vinicius Mikuni for useful discussions, and Gregor Kasieczka for suggesting performance comparison with the extended feature sets in Ref~\cite{finke2023BackToTheRoots}.  The work of TC is supported by the Fundamental Research Funds for the Central Universities and further sponsored by Shanghai Pujiang Programme. BN is supported by the U.S. Department of Energy (DOE), Office of Science under contract DE-AC02-76SF00515.  This research used resources of the National Energy Research Scientific Computing Center, a DOE Office of Science User Facility supported by the Office of Science of the U.S. Department of Energy under Contract No. DE-AC02-05CH11231.

\end{acknowledgments}

\bibliography{draft}

\newpage
\clearpage
\onecolumngrid

\section{Appendix}\label{sec:supplement}

For completeness, we summarize the standard observables and performance metrics used in this work in Table~\ref{tab:definitions}.

\begin{table}[htbp]
\centering
\footnotesize
\setlength{\tabcolsep}{5pt}
\renewcommand{\arraystretch}{1.15}
\begin{tabular}{l|p{0.66\linewidth}}
\hline\hline
\textbf{Quantity} & \textbf{Definition} \\
\hline
$\boldsymbol{m_J}$ 
& Invariant mass of a reconstructed jet, sensitive to its internal energy distribution. \\

$\boldsymbol{m_{JJ}}$ 
& Invariant mass of the two leading-$p_T$ jets, used to define the signal and sideband regions in the LHCO benchmark. \\

$\boldsymbol{\tau_{21}}$ 
& $N$-subjettiness ratio $\tau_2/\tau_1$, commonly used to identify two-prong jet substructure such as boosted $W/Z$ bosons. \\

$\boldsymbol{\tau_{32}}$ 
& $N$-subjettiness ratio $\tau_3/\tau_2$, commonly used to identify three-prong jet substructure such as boosted top quarks. \\

$\boldsymbol{\epsilon_s}$ 
& Signal efficiency: fraction of signal events passing a given classifier-score threshold. \\

$\boldsymbol{\epsilon_b}$ 
& Background efficiency: fraction of background events passing the same classifier-score threshold. \\

$\boldsymbol{\mathrm{SI}}$ 
& Significance improvement, defined as $\mathrm{SI} \equiv \epsilon_s/\sqrt{\epsilon_b}$, which approximates the improvement in statistical sensitivity relative to an inclusive selection. \\
\hline
\end{tabular}
\caption{Definitions of the standard observables and performance metrics used throughout the analysis.}
\label{tab:definitions}
\end{table}

Table~\ref{tab:ot_correlations} summarizes Pearson correlations and mutual information values between the leading OT PCA modes and standard jet-substructure observables, including jet masses and higher-order $N$-subjettiness ratios, computed on the same diagnostic sample used in Fig.~\ref{fig:pca_variance_plot}.

Interestingly, the leading OT feature (OT-PC1) shows a strong correlation with $m_{J2}$ in the R\&D1 dataset, despite the fact that the OT representation is constructed purely from jet morphology and contains no explicit mass information. This observation suggests that nontrivial correlations may exist between the geometric structures captured by OT and conventional jet observables, and warrants further investigation.

More generally, neither the correlations nor the mutual information values are saturated, indicating that the OT features are not simply redundant with the standard observables. This supports the interpretation that the OT representation captures overlapping but complementary information, providing a possible explanation for the performance gains observed when OT features are added to extended subjettiness feature sets.

\begin{table}[htbp]
\centering
\scriptsize
\setlength{\tabcolsep}{2.5pt}
\renewcommand{\arraystretch}{2}

\begin{minipage}{0.49\linewidth}
\centering
\begin{tabular}{l||rrrrr}
\hline\hline
$\boldsymbol{\rho}$ 
& \textbf{OT\boldmath{$_1$}} 
& \textbf{OT\boldmath{$_2$}} 
& \textbf{OT\boldmath{$_3$}} 
& \textbf{OT\boldmath{$_4$}} 
& \textbf{OT\boldmath{$_{100}$}} \\
\hline
$\boldsymbol{m_{J_1}}$          & $-0.342$ & $-0.188$ & $-0.628$ & $-0.035$ & $-0.016$ \\
$\boldsymbol{m_{J_2}}$          & $-0.911$ &  $0.073$ &  $0.279$ & $-0.143$ &  $0.002$ \\
$\boldsymbol{\tau_{21}^{J_1}}$  &  $0.337$ & $-0.061$ & $-0.013$ &  $0.206$ & $-0.004$ \\
$\boldsymbol{\tau_{21}^{J_2}}$  &  $0.677$ & $-0.197$ & $-0.286$ &  $0.536$ & $-0.007$ \\
$\boldsymbol{\tau_{4,1}^{J_1}}$ & $-0.214$ & $-0.165$ & $-0.516$ &  $0.027$ & $-0.005$ \\
$\boldsymbol{\tau_{4,2}^{J_1}}$ & $-0.109$ & $-0.167$ & $-0.477$ &  $0.083$ & $-0.010$ \\
$\boldsymbol{\tau_{4,5}^{J_1}}$ & $-0.262$ & $-0.121$ & $-0.391$ & $-0.035$ &  $0.002$ \\
$\boldsymbol{\tau_{4,1}^{J_2}}$ & $-0.375$ & $-0.008$ & $-0.017$ &  $0.455$ & $-0.013$ \\
$\boldsymbol{\tau_{4,2}^{J_2}}$ & $-0.264$ & $-0.060$ & $-0.044$ &  $0.459$ & $-0.010$ \\
$\boldsymbol{\tau_{4,5}^{J_2}}$ & $-0.396$ &  $0.039$ &  $0.002$ &  $0.369$ & $-0.010$ \\
$\boldsymbol{\tau_{9,1}^{J_1}}$ & $-0.184$ & $-0.175$ & $-0.529$ &  $0.048$ & $-0.006$ \\
$\boldsymbol{\tau_{9,1}^{J_2}}$ & $-0.376$ & $-0.015$ & $-0.026$ &  $0.443$ & $-0.021$ \\
\hline
\end{tabular}
\end{minipage}
\hfill
\begin{minipage}{0.49\linewidth}
\centering
\begin{tabular}{l||rrrrr}
\hline\hline
$\boldsymbol{I}$ 
& \textbf{OT\boldmath{$_1$}} 
& \textbf{OT\boldmath{$_2$}} 
& \textbf{OT\boldmath{$_3$}} 
& \textbf{OT\boldmath{$_4$}} 
& \textbf{OT\boldmath{$_{100}$}} \\
\hline
$\boldsymbol{m_{J_1}}$          & $0.198$ & $0.100$ & $0.253$ & $0.055$ & $0.025$ \\
$\boldsymbol{m_{J_2}}$          & $0.870$ & $0.333$ & $0.227$ & $0.284$ & $0.048$ \\
$\boldsymbol{\tau_{21}^{J_1}}$  & $0.087$ & $0.019$ & $0.022$ & $0.040$ & $0.000$ \\
$\boldsymbol{\tau_{21}^{J_2}}$  & $0.339$ & $0.211$ & $0.104$ & $0.309$ & $0.020$ \\
$\boldsymbol{\tau_{4,1}^{J_1}}$ & $0.082$ & $0.049$ & $0.160$ & $0.025$ & $0.008$ \\
$\boldsymbol{\tau_{4,2}^{J_1}}$ & $0.057$ & $0.019$ & $0.146$ & $0.015$ & $0.000$ \\
$\boldsymbol{\tau_{4,5}^{J_1}}$ & $0.086$ & $0.041$ & $0.102$ & $0.008$ & $0.011$ \\
$\boldsymbol{\tau_{4,1}^{J_2}}$ & $0.186$ & $0.040$ & $0.037$ & $0.191$ & $0.043$ \\
$\boldsymbol{\tau_{4,2}^{J_2}}$ & $0.111$ & $0.007$ & $0.035$ & $0.162$ & $0.028$ \\
$\boldsymbol{\tau_{4,5}^{J_2}}$ & $0.164$ & $0.049$ & $0.038$ & $0.147$ & $0.023$ \\
$\boldsymbol{\tau_{9,1}^{J_1}}$ & $0.067$ & $0.041$ & $0.185$ & $0.016$ & $0.001$ \\
$\boldsymbol{\tau_{9,1}^{J_2}}$ & $0.200$ & $0.043$ & $0.044$ & $0.185$ & $0.033$ \\
\hline
\end{tabular}
\end{minipage}

\caption{Pearson correlation coefficients $\rho(X,Y)$ (\textit{left}) and mutual information values $I(X;Y)$ (\textit{right}) between jet-substructure observables $X$ and OT-derived PCA features $Y$ for the R\&D1 diagnostic sample.}
\label{tab:ot_correlations}
\end{table}

\begin{figure*}[htbp]
    \centering
    \includegraphics[width=\linewidth]{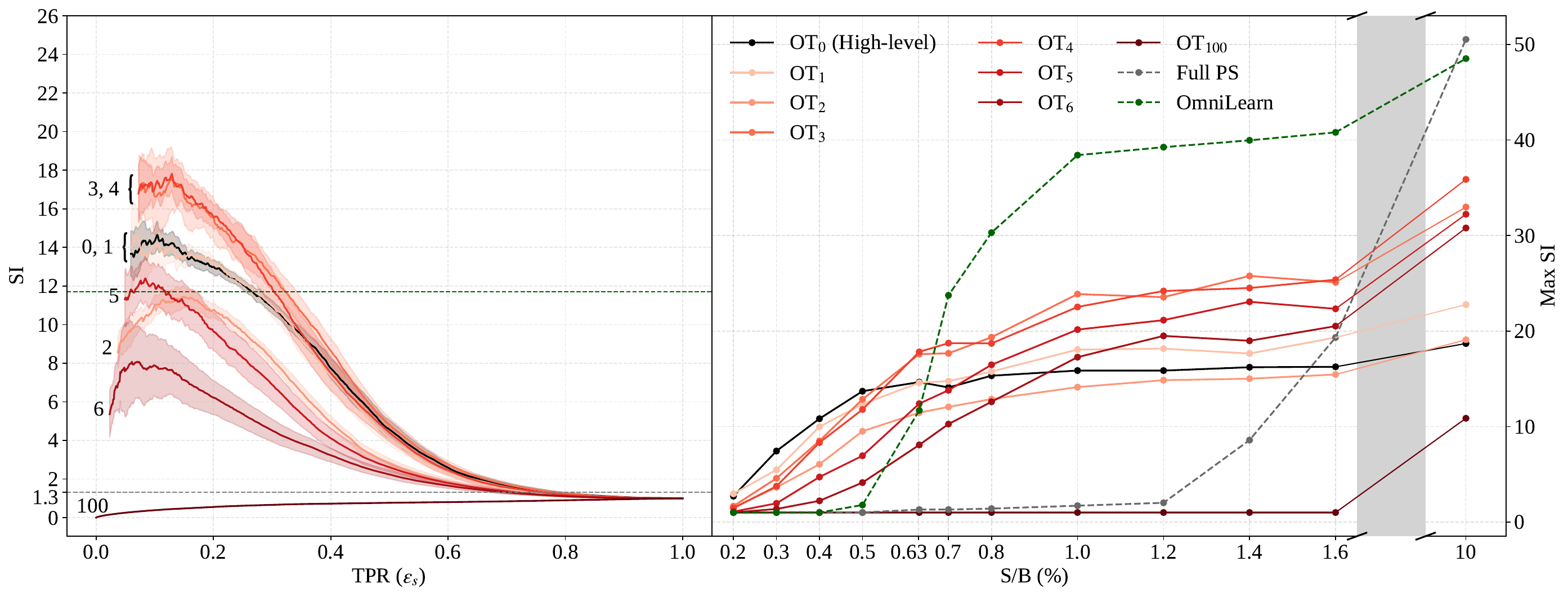}
    \caption{Anomaly detection for the R\&D1 dataset using MLP ensembles as the classifier. \textit{Left:} The SI curves at the signal injection level S/B$=0.63\%$. \textit{Right:} Maximum SI at S/B between 0.2\% and 10\%. The group of red curves represent increasing numbers of OT features added to standard high-level observables $\{m_{J_1}, m_{J_2}, \tau_{21}^{J_1}, \tau_{21}^{J_2} \}$, with shaded bands denoting $1\sigma$ variations across the ensembles of trained models. The dashed gray line shows the performance of the full phase space method from Ref.~\cite{Buhmann:2023acn}, whereas the dashed green line shows that of the pre-trained foundation model OmniLearn from Ref.~\cite{omnilearn}. Both the full phase space method and foundation models are evaluated in the same IAD setup.}
    \label{fig:vary_sig_R&D1}
\end{figure*}

\begin{figure*}[htbp]
    \includegraphics[width=\linewidth]{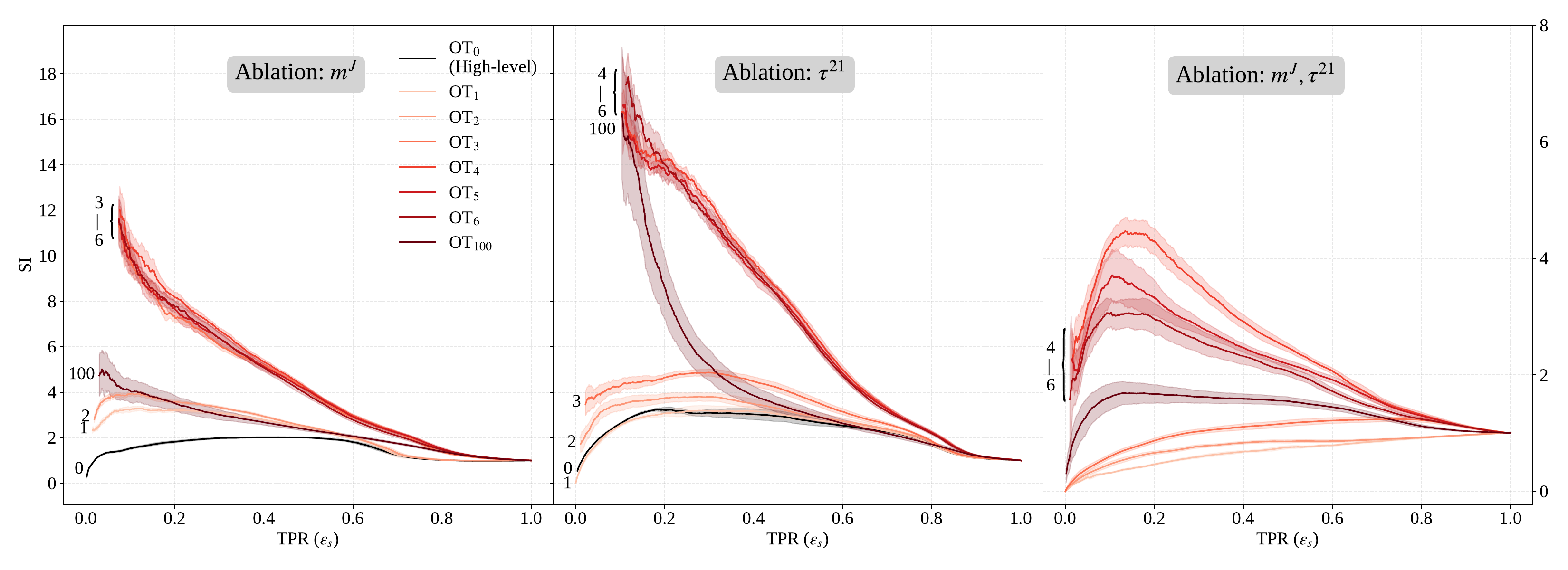}
    \caption{The SI curves for the R\&D1 dataset using BDT ensembles as the classifier at S/B$=0.63\%$. \textit{Left:} The input features are $\{\tau_{21}^{J_1}, \tau_{21}^{J_2} \}$ and different number of PCA modes for the OT representation (the group of red lines). \textit{Middle:} The input features are $\{m_{J_1}, m_{J_2} \}$ and different number of PCA modes for the OT representation. Note that the middle subplot shares the same $y$-axis with the leftmost subplot. \textit{Right:} The input features only include different number of PCA modes for the OT representation. Note that the rightmost subplot has its own $y$-axis to the right. Shaded bands denote $1\sigma$ variations across the ensembles of trained models.}
    \label{fig:ablation_R&D1}
\end{figure*}

\end{document}